\documentclass[manuscript]{aastex701}

\usepackage{graphicx}
\usepackage{enumerate}
\usepackage{xcolor}

\newcommand{\muvec}{\mbox{\boldmath $\mu$}}

\begin{document}
\title{KMT-2024-BLG-3237: Another Free-floating Planet Candidate with Angular Einstein Radius Measurement
}

\correspondingauthor{Youn Kil Jung}
\email{ykjung21@kasi.re.kr}

\author{Tanagodchaporn Inyanya}
\affiliation{Korea Astronomy and Space Science Institute, Daejeon 34055, Republic of Korea}
\affiliation{National University of Science and Technology (UST), Daejeon 34113, Republic of Korea}
\email{tinyanya@kasi.re.kr}

\author[0000-0002-0314-6000]{Youn Kil Jung}
\affiliation{Korea Astronomy and Space Science Institute, Daejeon 34055, Republic of Korea}
\affiliation{National University of Science and Technology (UST), Daejeon 34113, Republic of Korea}
\email{ykjung21@kasi.re.kr}

\author[0000-0003-0626-8465]{Hongjing Yang}
\affiliation{Department of Astronomy, Westlake University, Hangzhou 310030, Zhejiang Province, China}
\email{hongjing.yang@qq.com}

\author[0000-0002-9241-4117]{Kyu-Ha Hwang}
\affiliation{Korea Astronomy and Space Science Institute, Daejeon 34055, Republic of Korea}
\email{kyuha@kasi.re.kr}

\author{Andrew Gould} % No OrcID on purpose
\affiliation{Department of Astronomy, Ohio State University, 140 W. 18th Ave., Columbus, OH 43210, USA}
\email{gould.34@osu.edu}

\author[0000-0003-3316-4012]{Michael D. Albrow}
\affiliation{University of Canterbury, School of Physical and Chemical Sciences, Private Bag 4800, Christchurch 8020, New Zealand}
\email{michael.albrow@canterbury.ac.nz}

\author[0000-0001-6285-4528]{Sun-Ju Chung}
\affiliation{Korea Astronomy and Space Science Institute, Daejeon 34055, Republic of Korea}
\email{sjchung@kasi.re.kr}

\author[0000-0002-2641-9964]{Cheongho Han}
\affiliation{Department of Physics, Chungbuk National University, Cheongju 28644, Republic of Korea}
\email{cheongho@astroph.chungbuk.ac.kr}

\author[0000-0001-9823-2907]{Yoon-Hyun Ryu} 
\affiliation{Korea Astronomy and Space Science Institute, Daejeon 34055, Republic of Korea}
\email{yhryu@kasi.re.kr}

\author[0000-0002-4355-9838]{In-Gu Shin}
\affiliation{School of Science, Westlake University, Hangzhou, Zhejiang 310030, China}
\email{ingushin@gmail.com}

\author[0000-0003-1525-5041]{Yossi Shvartzvald}
\affiliation{Department of Particle Physics and Astrophysics, Weizmann Institute of Science, Rehovot 7610001, Israel}
\email{yossishv@gmail.com}

\author[0000-0001-9481-7123]{Jennifer C. Yee}
\affiliation{Center for Astrophysics $|$ Harvard \& Smithsonian, 60 Garden St.,Cambridge, MA 02138, USA}
\email{jyee@cfa.harvard.edu}

\author[0000-0001-6000-3463]{Weicheng Zang}
\affiliation{Department of Astronomy, Westlake University, Hangzhou 310030, Zhejiang Province, China}
\email{weicheng.zang@cfa.harvard.edu}

\author{Dong-Jin Kim}
\affiliation{Korea Astronomy and Space Science Institute, Daejeon 34055, Republic of Korea}
\email{keaton03@kasi.re.kr}

\author[0000-0003-0043-3925]{Chung-Uk Lee}
\affiliation{Korea Astronomy and Space Science Institute, Daejeon 34055, Republic of Korea}
\email{leecu@kasi.re.kr}

\author[0000-0002-6982-7722]{Byeong-Gon Park}
\affiliation{Korea Astronomy and Space Science Institute, Daejeon 34055, Republic of Korea}
\email{bgpark@kasi.re.kr}

\begin{abstract}
Planet formation theories suggest the presence of free-floating planets (FFPs) that are ejected from their formation sites. While these planets emit very little light, they can be identified through gravitational microlensing. Here, we report the discovery of a FFP candidate in the microlensing event KMT-2024-BLG-3237. The observed light curve exhibits strong finite-source effects characterized by a small amplitude $(\lesssim 0.9\,{\rm mag})$ and a short timescale $(\lesssim 3\,{\rm days})$. The analysis yields an Einstein timescale of $t_{\rm E} = 0.54\pm0.02\,{\rm days}$ and an angular Einstein radius of $\theta_{\rm E} = 6.30\pm0.48\,\mu{\rm as}$. The measurements make it possible to estimate the lens mass as $M \simeq 102\,M_{\oplus}\,(\pi_{\rm rel}/16\,\mu{\rm as})^{-1}$, where $\pi_{\rm rel}$ is the relative lens-source parallax. Depending on the unknown $\pi_{\rm rel}$, the lens could be a Neptune-mass planet $(\pi_{\rm rel} \simeq 0.1\,{\rm mas})$ or a Saturn-mass planet $(\pi_{\rm rel} \simeq 16\,\mu{\rm as})$. A Bayesian analysis yields the lens mass $M = {67.3}_{-42.5}^{+103.2}\,M_{\oplus}$ and the lens distance $D_{\rm L} = {7.34}_{-2.11}^{+0.96}\,{\rm kpc}$. This lens is the eleventh isolated microlens with a measurement of $\theta_{\rm E} < 10\,\mu{\rm as}$. We find that additional searches for possible signatures of a lens host do not show significant evidence for the host.
\end{abstract}

\section{Introduction} \label{sec:introduction}

Free-floating planets (FFPs) are substellar objects with planetary masses, which are not gravitationally attached to any host star. These substellar objects can be formed, like stars, through the collapse and fragmentation of gas clouds (e.g., \citealt{Luhman2012}). It is suggested that the minimum mass of the FFPs formed by this process ranges from $\sim 1$ to $\sim 10\,M_{\rm J}$ (e.g., \citealt{Whitworth2006, Luhman2012}). FFPs can also be formed in protoplanetary disks around stars, like ordinary planets, and then be ejected through various mechanisms. These include planet–planet scattering (e.g.,\citealt{RasioFord1996}), dynamical interactions with stars in multiple-star systems (e.g., \citealt{Kaib2013}), stellar flybys (e.g., \citealt{Malmberge2011}), or orbital disruption caused by the host star’s evolution beyond the main sequence phase (e.g., \citealt{VerasRaymond2012}).

Considering that different FFP formation scenarios predict different mass distributions of FFPs, understanding their origins requires a statistical characterization of the FFP population across a broad mass range, from terrestrial to Jovian masses. Many nearby high-mass $(M \gtrsim M_{\rm J})$ FFP candidates have been detected through high-resolution direct imaging (e.g., \citealt{PearsonMcCaughrean2023}). However, low-mass FFPs $(M < M_{\rm J})$ are rarely detected by the direct imaging method, as they remain below the brightness limits of currently available telescopes. In contrast, gravitational microlensing does not rely on the brightness of FFPs, and thus the microlensing method can provide a unique capability to detect FFPs down to terrestrial- (or even dwarf-planet-) mass objects \citep{Gould2021, Gould2024}.

The microlensing method observes the light from a distant source star, which is distorted and magnified by the gravitational field of a lensing object. A key observable in a microlensing event is the Einstein timescale, $t_{\rm E}$, which relies on the relative lens-source (parallax, proper motion), $(\pi_{\rm rel}, \mu_{\rm rel})$, and the angular Einstein radius, $\theta_{\rm E}$, i.e., 
\begin{equation}
    t_{\rm E} = \frac{\theta_{\rm E}}{\mu_{\rm rel}}, \quad
    \theta_{\rm E} \equiv \sqrt{\kappa M \pi_{\rm rel}}, \quad
    \pi_{\rm rel} = \mathrm{au} \left( \frac{1}{D_{\rm L}} - \frac{1}{D_{\rm S}} \right), 
    \label{eq:timescale}
\end{equation}
where $\kappa \equiv 4\,G / (c^2\,{\rm au}) =  8.14\,{\rm mas}\,M_{\odot}^{-1}$ and $M$ is the lens mass. Here, $D_{\rm L}$ and $D_{\rm S}$ are, respectively, the lens and the source distances. As $t_{\rm E} \propto M^{1/2}$, it is expected that Einstein timescales due to FFPs are very short:
\begin{equation}
    t_{\rm E} = \\
    0.8\, {\rm days} \left( \frac{M}{1\,M_{\rm J}} \right)^{1/2}
    \left( \frac{\pi_{\rm rel}}{16\,\mu {\rm as}} \right)^{1/2}
    \left( \frac{\mu_{\rm rel}}{5\,{\rm mas}\,{\rm yr}^{-1}} \right)^{-1}.
    \label{eq:mass_pirel_urel}
\end{equation}
We expect $t_{\rm E} \simeq 0.8\,{\rm days}$ for a Jupiter-mass lens, assuming typical values of $\pi_{\rm rel} = 16\, {\mu}{\rm as}$ and $\mu_{\rm rel} = 5$ ${\rm mas}$ ${\rm yr}^{-1}$, which correspond to typical events generated by lenses in the Galactic bulge. However, a secure lens-mass determination requires the measurements of both $\mu_{\rm rel}$ and $\pi_{\rm rel}$.

In the cases for which the lens closely approaches or transits the source, the source size affects the microlensing light curve \citep{Gould1994,Nemiroff&Wickramasinghe1994,Witt&Mao1994}. For microlensing events that exhibit such finite-source effects, one can measure $\theta_{\rm E}$ based on the detection of the normalized source radius $\rho=\theta_{*} / \theta_{\rm E}$. Here, $\theta_{*}$ is the angular source radius. The $\theta_{\rm E}$ measurement then enables us to estimate the lens mass with better accuracy, as it partially resolves the three-way degeneracy in $t_{\rm E}$ by eliminating $\mu_{\rm rel}$ as an unknown parameter:
\begin{equation}
    M = \frac{\theta_{\rm E}^2}{\kappa \pi_{\rm rel}} = \\
    0.16\,M_{\rm J} \left( \frac{\theta_{\rm E}} {5\,\mu {\rm as}} \right)^2 \\
    \left( \frac{\pi_{\rm rel}}{16\,\mu {\rm as}} \right)^{-1}.
    \label{eq:mass_thetae_pirel}
\end{equation}
Finite-source effects can be strong for planetary-lens events, as their Einstein radii are comparable to the angular radii of sources, i.e., $\rho \approx 1$ \citep{BennettRhie1996}. In addition, if $\rho > 1$, strong finite-source effects increase the duration of events, which makes it possible to take more observations at a fixed cadence. This makes short-duration finite-source/point-lens (FSPL) events on giant-source stars easier to detect than those on dwarf-source stars or short-duration point-source/point-lens (PSPL) events. Hence, one would expect that an efficient way to identify FFP candidates is to observe giant-source microlensing events \citep{Kim2021,Gould2022,Koshimoto2023}.

To resolve the $(M, \pi_{\rm rel}, \mu_{\rm rel})$ degeneracy requires one additional observable. This is the microlens parallax $\pi_{\rm E}$, which is related to the lens mass and the relative parallax by \citep{Gould2000}   
\begin{equation}
    M = {\theta_{\rm E} \over {\kappa\pi_{\rm E}}} ; \quad \pi_{\rm E} = {\pi_{\rm rel} \over \theta_{\rm E}}.
    \label{eq:microlensparallax}
\end{equation}
The microlens parallax can be measured using the orbital acceleration of Earth \citep{Gould1992}. However, this approach is challenging for FFP events due to their short timescales \citep{Gould&Yee2013}. The alternative way to measure the parallax is to use simultaneous observations from two widely separated (ground- and space-based) observatories \citep{Refsdal1966}. For example, the $Spitzer$ telescope carried out microlensing observations with ground-based telescopes during the $Spitzer$ microlensing campaign \citep{Gould2013Spitzer,Gould2014Spitzer,Gould2015aSpitzer,Gould2015bSpitzer,Gould2016Spitzer,Gould2018Spitzer}. These observations made it possible to detect microlens parallaxes for a large number of single- and binary-lens events, including bound-planet systems (e.g., \citealt{Udalski2015,Yee2015,Shvartzval2017,Gould2020}). The $Kepler$ observations from the K2 Campaign 9 survey also provided $\pi_{\rm E}$ measurements for several events (e.g., \citealt{Henderson2016,Zhu2017,Zang2018}). Although parallax measurements of FFP events from these surveys were not reported, \citet{Ban2020} showed that based on space-based parallax simulations, ground- and space-based telescopes could enable parallax measurements for FFPs down to $\sim 0.5\,M_{\oplus}$ mass objects.

To date, ten short-timescale FSPL events have been reported \citep{Mroz2018,Mroz2019,Mroz2020a,Mroz2020b,Kim2021,Ryu2021,Koshimoto2023,Jung2024,Kapusta2025}. All of these are characterized by $t_{\rm E} < 2\,{\rm days}$ and $\theta_{\rm E} < 10\,\mu {\rm as}$, the latter of which is a useful threshold for planetary-mass lenses \citep{Gould2022}. These measurements imply that the lenses are either FFPs or wide-orbit $(\gtrsim 10\,{\rm au})$ planets, as microlensing observations alone cannot exclude the possibility of a distant host star (e.g., \citealt{Clanton&Gaudi2017}). If these planetary-mass lenses are real FFPs, statistical studies \citep{Gould2022,Sumi2023} suggest that unbound planets are more common than wide-orbit planets and/or stellar objects. \citet{Yee2025} have shown that, regardless of whether they are bound or free-floating, this population places interesting constraints on wide-orbit planet populations.

Here, we present the discovery of another FFP candidate identified in the microlensing event KMT-2024-BLG-3237, with measurements of $t_{\rm E} \simeq 0.54$ days and $\theta_{\rm E} \simeq 6.3\,\mu\mathrm{as}$. The lens magnified a giant source star with $\theta_{\rm *} \simeq 6.42\,\mu\mathrm{as}$, located in the Galactic bulge.

\section{Observations} \label{sec:observation}

KMT-2024-BLG-3237 occurred at $({\rm R.A.}, {\rm decl.})_{\rm J2000} = $(17:50:31.23, $-32$:15:48.89) $[(l, b) = (-2 \fdg 31, -2 \fdg 62)]$. This event was discovered by the Korea Microlensing Telescope Network (KMTNet; \citealt{Kim2016}), which consists of three $1.6\,{\rm m}$ telescopes equipped with $2^{\circ} \times 2^{\circ}$ cameras. The KMTNet telescopes are distributed at the South African Astronomical Observatory (KMTS), the Cerro Tololo Inter-American Observatory (KMTC), and the Siding Spring Observatory (KMTA). The event was observed in the overlap region of KMTNet BLG41 and BLG22 fields, with cadences of $\Gamma = 2\,{\rm hr}^{-1}$ for BLG41 and $\Gamma = 1\,{\rm hr}^{-1}$ for BLG22, which combine to a total cadence of $\Gamma = 3\,{\rm hr}^{-1}$. KMTNet observations were mainly taken in the Cousins $I$ band, but after every 10th $I$-band observation, there was one observation in the Johnson $V$ band. These observations were initially processed using the pySIS pipeline \citep{Albrow2009}, which was developed based on the Difference Image Analysis (DIA; \citealt{Tomany&Crotts1996}; \citealt{Alard&Lupton1998}).

KMT-2024-BLG-3237 was first identified in 2025 February through the KMTNet EventFinder package \citep{Kim2018}. The event, which showed deviations that would be indicative of FSPL events, was immediately recognized as an FFP candidate during the inspection of events found by the package. For the light-curve analysis, we then reprocessed the data sets using the refined pySIS pipeline \citep{Yang2024}.

Microlensing experience shows that the error bars measured from photometry pipelines are monotonically connected to the true errors, but the transformation from measured to true errors varies between events and/or between observatories. To address this effect, we rescale the error bars based on the procedure presented in \citet{Yee2012}. We use the formula
\begin{equation}
\sigma^\prime = k \sqrt{\sigma_{0}^2 + (e_{\rm min})^2},
\label{eq:error}
\end{equation}
where $(k, e_{\rm min})$ are the scaling parameters and $\sigma_{0}$ is the error measured from the pipeline. For each data set, we choose $e_{\rm min}$ to make the cumulative $\chi^{2}$ distribution ordered by source magnification linear. We next rescale the errors using $k$ to make $\chi^{2}/{\rm dof}$ unity. We note that this renormalization process is carried out based on the FSPL model.

\section{Light Curve Analysis} \label{sec:analysis}

\begin{figure}
  \centering
  \includegraphics[width=0.8\linewidth]{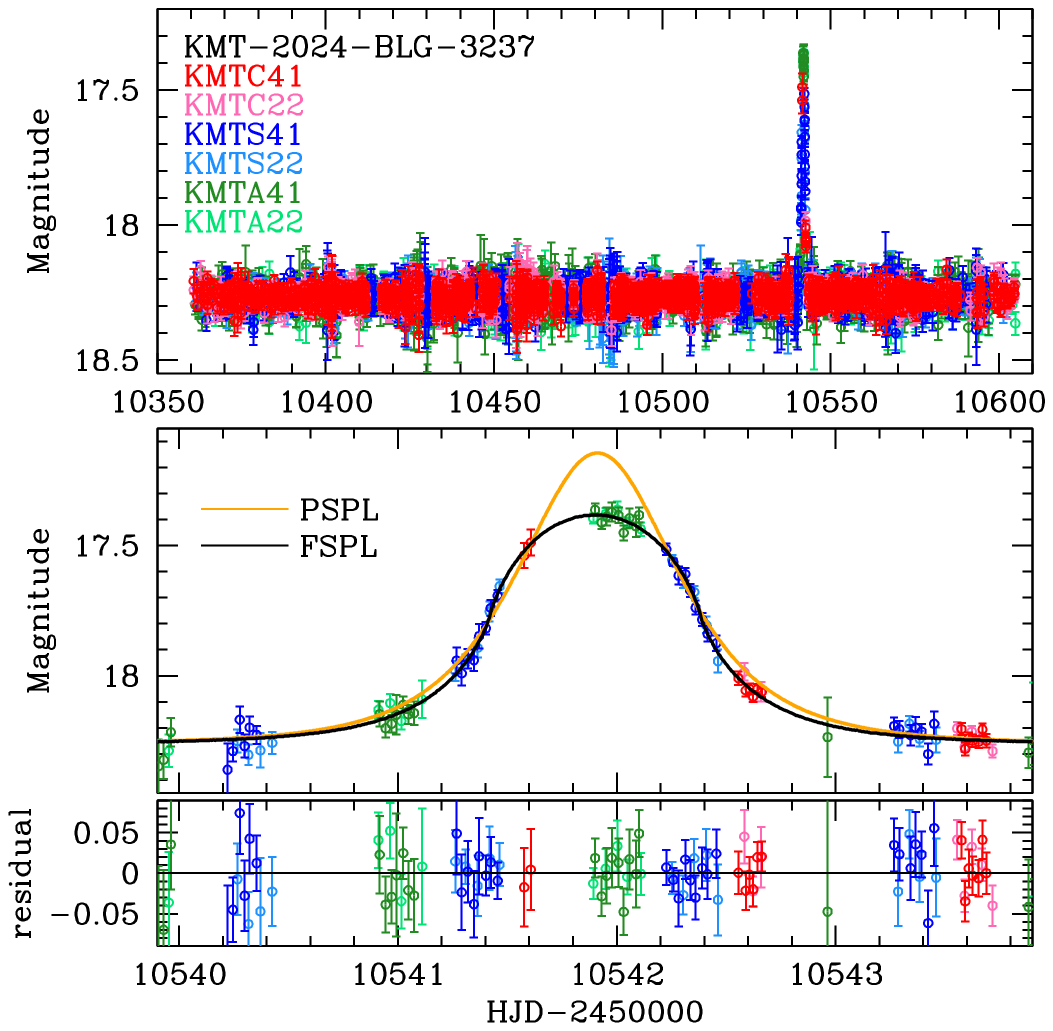}
  \caption{Light curve of KMT-2024-BLG-3237. The upper panel displays the full light curve from the 2024 season, while the lower panel provides the zoom-in of the event. The black and orange curves indicate the FSPL $(f_{\rm B} = 0)$ and the PSPL model, respectively.}
  \label{fig:lightcurve}
\end{figure}

As shown in Figure \ref{fig:lightcurve}, the observed light curve exhibits deviations from a standard PSPL model, which is described by three geometric parameters $(t_0, u_0, t_{\rm E})$. Here, $t_0$ is the time of maximum magnification and $u_0$ is the lens-source separation at $t_0$ (normalized to $\theta_{\rm E}$), i.e., the impact parameter. Such deviations can be described by the FSPL model, which requires an additional fitting parameter $\rho$. For each observatory, we consider two flux parameters, $(f_{\rm S}, f_{\rm B})$, to describe the source flux and any blended flux, respectively. The observed flux $F(t)$ is then given by
\begin{equation}
    F(t) = f_{\rm S}\,A(t) + f_{\rm B},
    \label{eq:observedflux}
\end{equation}
where $A(t)$ is the time-dependent lensing magnification. With these parameters, we fit the FSPL model to the data using the Markov Chain Monte Carlo (MCMC) technique. In this modeling, we utilize the inverse ray shooting method to calculate the finite-source magnification. To describe the source surface brightness profile, we also consider a linear limb-darkening model, i.e.,
\begin{equation}
    \frac{S_\lambda}{\overline{S_\lambda}} = 1 - \Gamma_{\lambda}(1-\frac{3}{2}\cos\theta),
    \label{eq:limb-darkening}
\end{equation}
where $\overline{S_\lambda} = F_{{\rm S}, \lambda}/\pi{\theta_*}^2$ is the mean source surface brightness, $F_{{\rm S}, \lambda}$ is the total source flux, $\Gamma_\lambda$ is the limb-darkening coefficient, $\lambda$ is the passband, and $\theta$ is the angular position on the source from its center. To find the coefficient $\Gamma_{\lambda}$, we use the linear limb-darkening model of \citet{Claret2000}, which is based on the intensity profile of a star, and use the relation $\Gamma_\lambda = 2u_{\lambda}/(3-u_{\lambda})$ \citep{Albrow2001}, where $u_{\lambda}$ is the limb-darkening coefficient of \citet{Claret2000} model. We adopt $\Gamma_{I} = 0.53$ based on the clump giant source estimated in Section~\ref{sec:physicalparameters}.

As listed in Table \ref{tab:FSPL}, we test two FSPL models: one with $f_{\rm B}$ as a variable and the other with $f_{\rm B}$ held fixed at zero $(f_{\rm B} = 0)$. We note that we adopt the KMTS BLG41 (KMTS41) as the representative observatory for the flux measurements, because this observatory densely covers the ascending and descending parts of the light curve, which enables us to precisely constrain the flux parameters. We find that both models give consistent results within $1\sigma$, with a minor $\chi^{2}$ difference of $\Delta\chi^2 = 0.7$. That is, the $\chi^2$ difference is in the range of statistical noise. As discussed by \citet{Park2004}, the negative blending in the free $f_{\rm B}$ model can be caused by low-level unseen systematics and/or statistical noise in the data. In addition, \citet{Mroz2019_OGLE4_OpticalDepth_EventRate} discussed that the distribution of blending fractions, $\eta = f_{\rm B}/(f_{\rm S}+f_{\rm B})$, of red giant stars is bimodal with values centering near $0$ and $1$. This implies that the observed flux is either dominated by the source $(\eta \sim 0)$ or by the blended light $(\eta \sim 1)$. Because our results are consistent with $\eta \sim 0$ and because the source position measured from difference images is closely aligned to the baseline object position $[\Delta(x, y) = (x, y)_{\rm S} - (x, y)_{\rm base} = (-0.007, 0.011)\,{\rm pixels} \simeq (-2.8, 4.4)\,{\rm mas}]$, we adopt the $f_{\rm B} = 0$ model as the preferred solution.

\section{Physical Parameters} \label{sec:physicalparameters}

\subsection{Source Star} \label{sec:sourcecolor}
The $\rho$ measurement enables us to infer $\theta_{\rm E} = \theta_{*}/\rho$. Typically, one can measure $\theta_*$ based on the source flux measurements in two different passbands within the same photometric system. However, this method is not viable for the current case because we lack $V$-band images taken during the event. As an alternative, we adopt the color and magnitude of the baseline object as those of the source because our results show the absence of evidence for the blended flux (see Table \ref{tab:FSPL}) and because the $I$-band source magnitude $I_{\rm S} = 18.26$ (inferred from the fitted source flux $f_{\rm S} = 0.785$) is consistent with that of the baseline object $I_{\rm base} = 18.26 \pm 0.02$. Here, the source magnitude is given by $I_{\rm S} = 18 - 2.5\,{\rm log}\,f_{\rm S}$.

\begin{figure}
  \centering
  \includegraphics[width=0.9\linewidth]{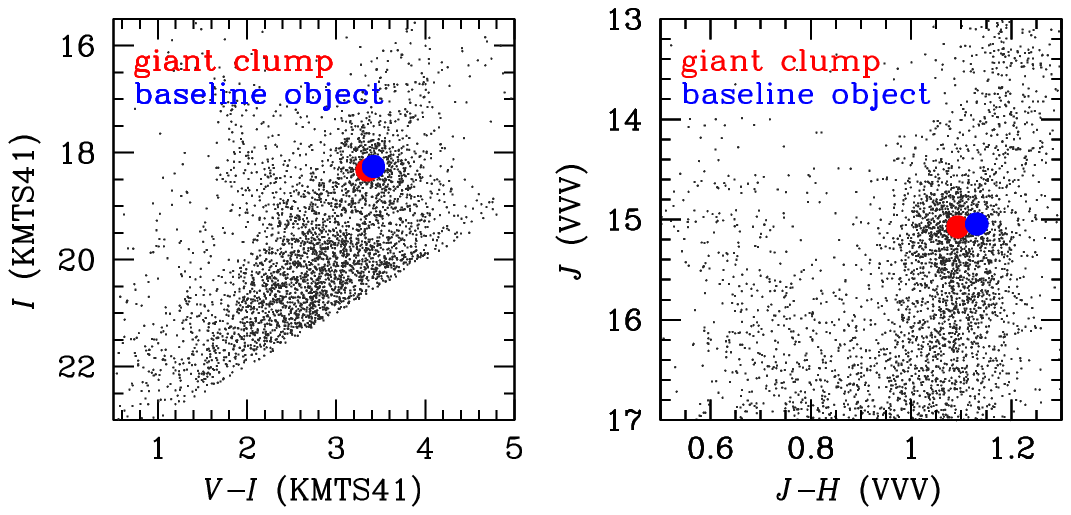}
  \caption{KMTS41 and VVV CMDs for stars in the vicinity (within $4^{\prime}$) of KMT-2024-BLG-3237. In each panel, the giant clump and baseline object (source) positions are marked by the red and blue dots, respectively.}
  \label{fig:CMD}
\end{figure}

We estimate $\theta_{*}$ using the standard method of \citet{Yoo2004}. We first make the color-magnitude diagram (CMD) with the KMTS41 observations, which includes nearby stars within $4^{\prime}$ of the event (see Figure \ref{fig:CMD}). In this CMD, the baseline object position is $(V-I, I)_{\rm base} = (3.42 \pm 0.09, 18.26 \pm 0.02)$. As described above, we adopt this measurement as the source position: $(V-I, I)_{\rm S} = (V-I, I)_{\rm base}$. We also find the giant clump (GC) position as $(V-I, I)_{\rm GC} = (3.35 \pm 0.02, 18.33 \pm 0.02)$. The source offset from the GC is then $\Delta(V-I, I) = (0.07 \pm 0.09, -0.07 \pm 0.02)$. We next adopt the dereddened GC position as $(V-I, I)_{\rm GC,0} = (1.06, 14.56)$ from the measurements of \citet{Nataf2013} and \citet{Bensby2013}, respectively. The dereddened source position is then
\begin{equation}
    \label{eq:cmdoffset}
    (V-I, I)_{\rm S,0} = \Delta(V-I, I) + (V-I, I)_{\rm GC,0} = (1.13 \pm 0.10, 14.50 \pm 0.03), 
\end{equation}
which indicates that the lensed source is a K1- or K2-type red giant.

Due to heavy extinction along the line of sight, however, the uncertainty in the $(V-I)_{\rm S}$ measurement is relatively large. To find a more precise estimate for the source color, we conduct an additional $(J-H, J)$ CMD analysis using the catalogue from the VISTA Variables in the Via Lactea (VVV) survey \citep{Minniti2023}. From this, we find $(J-H, J)_{\rm S} = (1.13 \pm 0.04, 15.04 \pm 0.03)$ and $(J-H, J)_{\rm GC} = (1.09 \pm 0.01, 15.07 \pm 0.01)$ (see Figure \ref{fig:CMD}). The source offset from the GC is $\Delta(J-H, J) = (0.04 \pm 0.04, -0.03 \pm 0.03)$. We next convert $(V-I, I)_{\rm GC,0} = (1.06, 14.56)$ to $(J-H, J)_{\rm GC,0} = (0.57, 13.84)$ using the \citet{Bessell&Brett1988} color-color relations. The dereddened source position is then $(J-H, J)_{\rm S,0} = (0.61 \pm 0.04 , 13.81 \pm 0.03)$. Finally, we convert $(J-H, J)_{\rm S,0}$ to $(V-I, I)_{\rm S,0}$ using the same color-color relations, which yields $(V-I,I)_{\rm S,0} = (1.13 \pm 0.04 , 14.49 \pm 0.03)$. We find that the KMTS41 and VVV source color measurements are consistent with each other, but the latter is more tightly constrained. Therefore, we use the VVV color to determine $\theta_{*}$.

We then convert $(V-I,I)_{\rm S,0}$ to $(V-K, K)_{\rm S,0}$ using again the \citet{Bessell&Brett1988} relations. Based on the the color-surface brightness relations from \citet{Kervella2004}, we subsequently estimate the angular source radius as  
\begin{equation}
    \theta_* = 6.42 \pm 0.45\, \mu {\rm as},
\end{equation}
where we add a quadratic $5\%$ error to the $\theta_*$ measurement to account for systematic errors and our assumption that $f_{\rm B} = 0$. We find
\begin{equation}
    \theta_{\rm E} = 6.30 \pm 0.48\, \mu {\rm as},~~~~~ \mu_{\rm rel} = 4.29 \pm 0.33\, {\rm mas}\,{\rm yr}^{-1}.
\end{equation}
It is worthwhile to note that the source color is estimated from the baseline object based on our assumption that $f_{\rm B} = 0$. If there is substantial blending (contrary to our assumption), the color could be incorrectly estimated, which in turn would affect the $\theta_{\rm E}$ measurement.

\subsection{Proper motion of the Source} \label{sec:sourcepropermotion}

\begin{figure}
  \centering
  \includegraphics[width=0.8\linewidth]{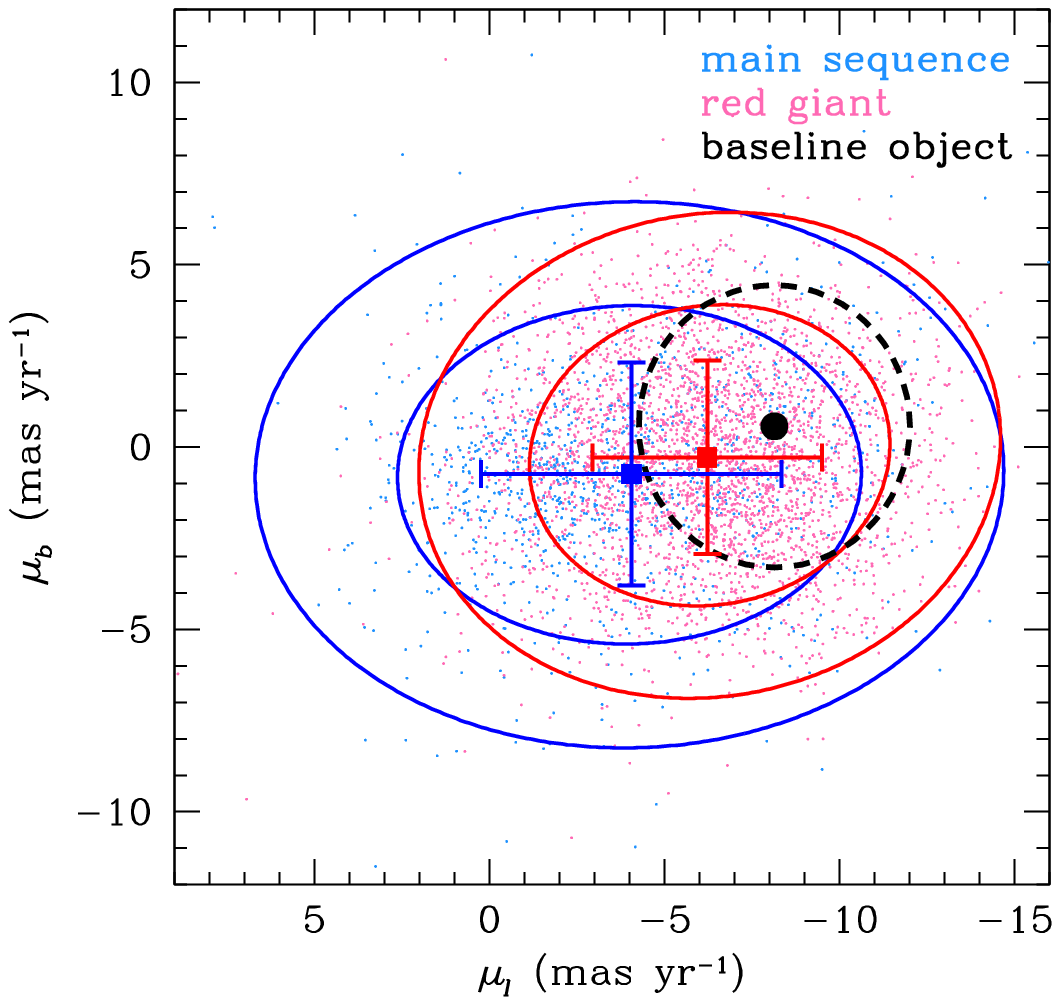}
  \caption{Gaia proper motions of stars in the vicinity (within $6^{\prime}$) of KMT-2024-BLG-3237. The red dots are red giant (bulge population) stars, while the blue dots are main sequence (disk population) stars. In each population, the inner and outer contours envelope 68$\%$ and 95$\%$ of the stars, respectively. The red square corresponds to the mean proper motion of the bulge population, $<\muvec_{\rm bulge}> (l,b) = (-6.22, -0.29) \pm (3.28, 2.65) \,{\rm mas}\,{\rm yr^{-1}}$. The blue square is the mean proper motion of the disk population, $<\muvec_{\rm disk}> (l,b) = (-4.06, -0.74) \pm (4.29, 3.06) \,{\rm mas}\,{\rm yr^{-1}}$. The baseline object (source) is located at the black dot. The black circle represents the locus of the relative lens-source proper motions, $\mu_{\rm rel} = 4.29\,{\rm mas}\,{\rm yr^{-1}}$.}
  \label{fig:GaiaPM}
\end{figure}

Our results show that the source is bright (as estimated above) and the blending associated with the event is negligible. In this case, one can adopt the proper motion of the baseline object derived by Gaia as that of the source (e.g., \citealt{Jung2024}). Figure~\ref{fig:GaiaPM} shows proper motions of stars in the vicinity of the event (within $6^{\prime}$) from the third Gaia data release (DR3; \citealt{GaiaCollaboration2023}). We rotate the Gaia measurements to Galactic coordinates using the coordinate transformation from \citet{Bachelet2018}. The red and blue dots represent red giant (bulge population) and main sequence (disk population) stars, respectively\footnote{We adopt the Gaia DR3 stars that satisfy the $RP$ magnitude of $ 19.2 < RP < 12.0$ and the $(BP - RP)$ color of $0.8 < (BP - RP) < 2.8$ as the main-sequence stars, while we adopt the DR3 stars with $19.1 < RP < 13.5$ and $2.9 < (BP - RP) < 4.3$ as the red giant stars. In each population, we exclude the stars that have the RUWE measurements of ${\rm RUWE} > 1.5.$}. The source proper motion is then
\begin{equation}
    \label{eq:gaiasource}
    \muvec_{\rm S}(l, b) = (-8.14, 0.57) \pm (0.44, 0.51)\,{\rm mas}\,{\rm yr}^{-1}.
\end{equation}
Here, Gaia reports the renormalized unit weight error (RUWE) value for this measurement as ${\rm RUWE} = 1.036$, which indicates that the measurement errors are consistent with the excess scatter of the data for the 5-parameter astrometric solution. That is, astrometric systematic effects have no significant impact on the measurement. From the $\mu_{\rm rel}$ and $\muvec_{\rm S}$ measurements, we find that the lens proper motion $\muvec_{\rm L} = \muvec_{\rm S} + \muvec_{\rm rel}$ lies close to the black circle in Figure~\ref{fig:GaiaPM}. This implies that $\muvec_{\rm L}$ is compatible with either bulge or disk lenses.

\subsection{Constraints on the host star} \label{sec:hostconstraints}

The light-curve analysis, including the measurements of $t_{\rm E} \simeq 0.54\,{\rm days}$ and $\theta_{\rm E} \simeq 6.30\,\mu{\rm as}$, suggests that the lens of KMT-2024-BLG-3237 is an FFP candidate. However, it is also possible that the lens orbits a widely separated host star. In such a case, the host may produce weak anomalies in the light curve. We therefore investigate the event with the binary-lens/single-source (2L1S) interpretation.

For the 2L1S analysis, we follow the method described in \citet{Kim2021}. We extend the FSPL model to the 2L1S case by introducing three binary-lens parameters $(s, q, \alpha)$, i.e., the projected planet-host separation, their mass ratio, and the angle between the direction of $\muvec_{\rm rel}$ and the planet-host axis (source trajectory angle). Here, the parameters $(u_{0}, t_{\rm E}, s, \rho)$ are scaled to the Einstein radius corresponding to the binary lens. We adopt a coordinate system centered on the caustic generated by the planet (planetary caustic) in order that $t_{0}$ remains comparable to the FSPL solution listed in Table \ref{tab:FSPL}. Because the binary-lens parameter space is non-linear, we first conducted a grid search over fixed $(s,q)$ grids with several trial values of $\alpha$, which are uniformly distributed on the unit circle. From this grid search, we identify two local minima. We refine these minima with additional MCMC runs, but this time with all fitting parameters are allowed to vary.

We find no 2L1S solutions that provide significant $\chi^{2}$ improvement over the FSPL solution. As listed in Table~\ref{tab:2L1S}, we find two (Local A and B) 2L1S solutions. However, the measured $(s, q, \alpha)$ are weakly constrained for both solutions, and their $\chi^{2}$ difference is just $\Delta\chi^{2} = 5.3$. Compared to the adopted FSPL $(f_{\rm B} = 0)$ model, the best-fit Local B solution gives a low-level $\chi^{2}$ improvement of $\Delta\chi^{2} = 6.4$ with four additional degrees of freedom. Assuming Gaussian statistics, this improvement corresponds to a significance of $p = (1+\Delta \chi^2/2)\exp(-\Delta \chi^{2}/2) = 18\%$. Similar to the FSPL solutions, the Local B solution exhibits the negative blending, and thus we additionally test the solution with $f_{\rm B} = 0$. From this, we find that the two Local B solutions give consistent results within $1\sigma$ and their $\chi^{2}$ difference is just $\Delta\chi^2 = 0.7$. For the Local A solution, the source encloses both the central and the planetary caustics, and the strong finite-source effects suppress the caustic signatures. In addition, this solution improves the fit by just $\Delta\chi^{2} = 1.1$ over the adopted FSPL model. These imply that the Local A solution is basically identical to the FSPL model except with an additional object.

Given that the event shows no significant signals that would be indicative of the lens host, we place a constraint on the lower limit of the projected planet-host separation. Following the method of \citet{Gaudi2000}, we build a grid in the $(q, s, \alpha)$ space, which consists of $61\times101\times37$ points that are uniformly distributed over $0 \leq {\rm log}\,q \leq 6$, $0 \leq {\rm log}\,s \leq 2$, and $0 \leq \alpha \leq 2\pi$. At each point, we fit the 2L1S model to the event using the MCMC with $(q, s, \alpha)$ held fixed and obtain the goodness of the fit $\chi^2_{\rm 2L1S}$ to find the $\chi^2$ difference between the model and the adopted FSPL solution, i.e., $\Delta\chi^{2} = \chi^{2}_{\rm 2L1S} - \chi^{2}_{\rm FSPL}$. For each $(q, s)$ pair, we then derive the detection probability of the host from the fraction of 2L1S models that satisfy $\Delta\chi^{2} < \Delta\chi^{2}_{\rm thresh}$. Here, we set the threshold value as $\Delta\chi^{2}_{\rm thresh} = 50$, given that the $\chi^{2}$ improvement of the Local B model relative to the FSPL solution is just $\Delta\chi^2 = 6.4$ (see Tables~\ref{tab:FSPL} and~\ref{tab:2L1S}).

\begin{figure}
  \centering
  \includegraphics[width=0.8\linewidth]{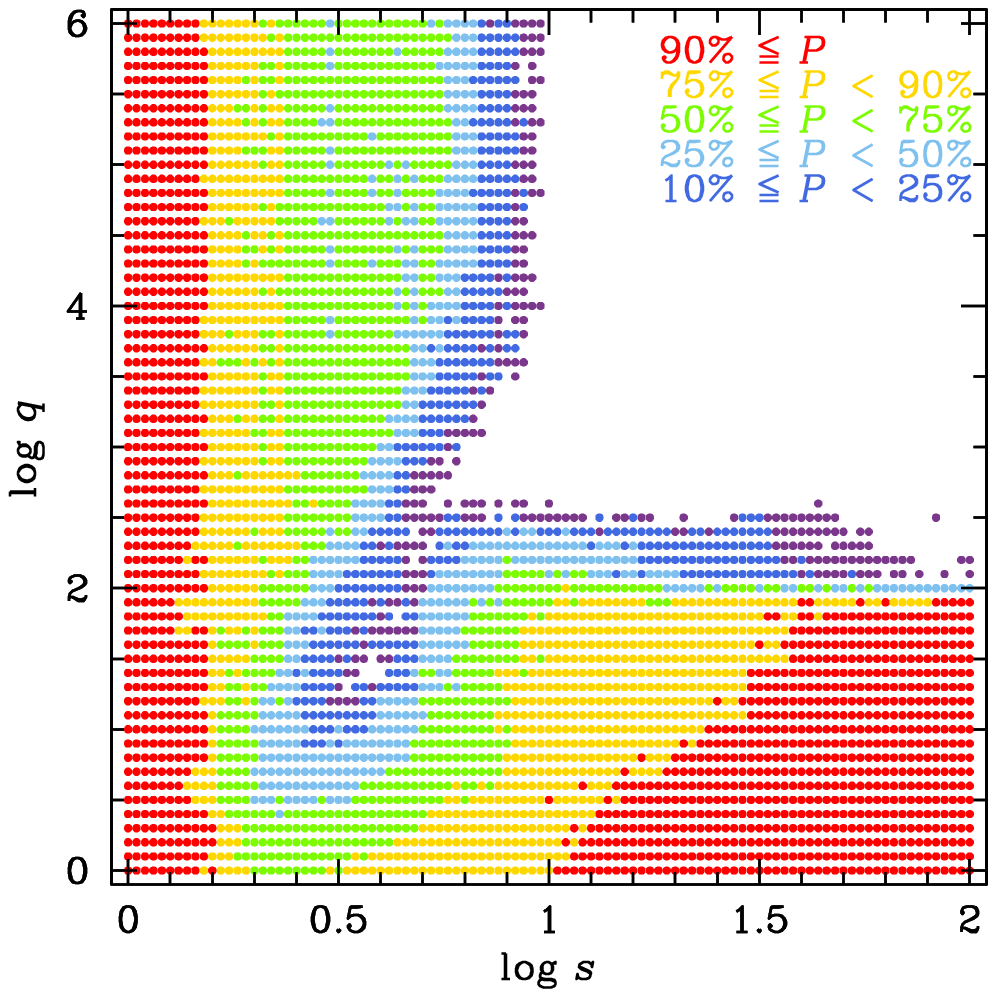}
  \caption{Lens host detection probability in the $(q, s)$ plane.}
  \label{fig:detectionprobability}
\end{figure}

We find that for putative hosts with ${\rm log}\,{q} > 1.9$, the $90\%$ lower separation limit corresponds to $s = 1.8$ (see Figure~\ref{fig:detectionprobability}). If we assume a $0.6\,M_{\odot}$ host with $\pi_{\rm rel} = 0.1\,{\rm mas}$ (disk lens), the lower limit for the physical projected separation would be $a_{\bot} = s{\theta_{\rm E,host}}{D_{\rm L}} \sim 6\,{\rm au}$. If we assume a $ 0.6\,M_{\odot}$ host with $\pi_{\rm rel} = 16\,\mu{\rm as}$ (bulge lens), the lower limit would be $a_{\bot} \sim 4\,{\rm au}$.

\section{Discussion} \label{sec:discussion}

\begin{figure}
  \centering
  \includegraphics[width=0.9\linewidth]{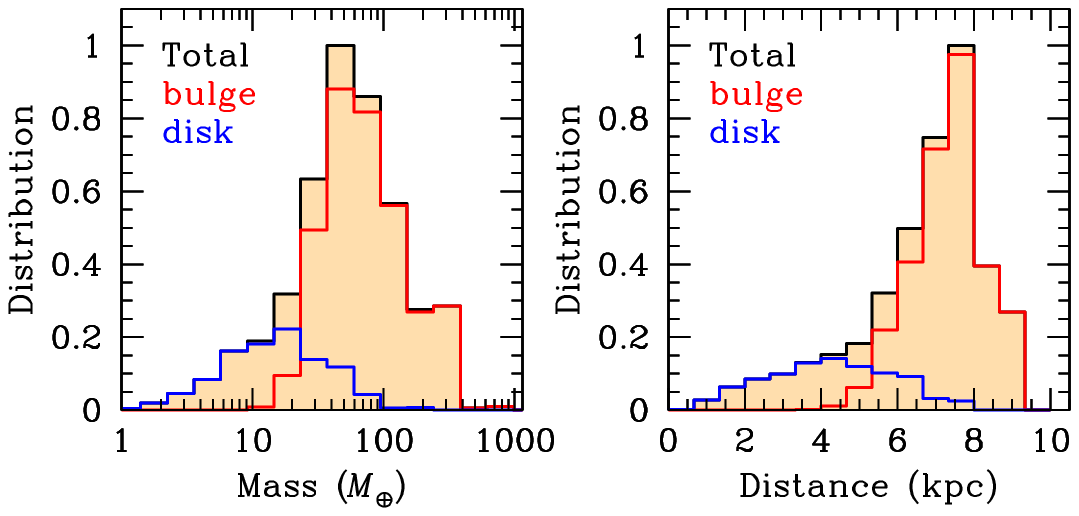}
  \caption{Posterior distributions of $M$ (left) and $D_{\rm L}$ (right) for KMT-2024-BLG-3237. In each panel, the blue and red distributions correspond to the disk- and bulge-lens contributions, respectively. The black distribution shows the total contribution.}  
  \label{fig:bayesian}
\end{figure}

KMT-2024-BLG-3237 is well explained by an FSPL model, which yields $t_{\rm E} = 0.54\pm0.02\,{\rm days}$ and $\theta_{\rm E} = 6.30\pm0.48\,\mu{\rm as}$. The measurements suggest that the lensing object is an FFP candidate, even though the lens mass is not unambiguously derived:
\begin{equation}
\label{eq:lensmass}
    M = 102\,M_{\oplus} \left({16\,\mu{\rm as} \over \pi_{\rm rel}} \right).
\end{equation}
The lens could be a Neptune- to Saturn-mass planet, depending on the unknown $\pi_{\rm rel}$. In principle, one can constrain the lens mass with the Bayesian approach, which is based on Galactic priors including mass functions (MFs), density profiles (DPs), and velocity distributions (VDs). For the MF, we adopt the 4-component power law model from \citet{Sumi2023}, which incorporates planetary, substellar, and stellar components. For the DP and VD, we use the models presented in \citet{Jung2021}. We then perform a Bayesian analysis with the measured constraints $(t_{\rm E}, \theta_{\rm E})$, following the procedure presented in \citet{Jung2018}. The posterior distributions for $M$ and $D_{\rm L}$ are shown in Figure~\ref{fig:bayesian}. The median and $68\%$ confidence intervals of the distributions are then  
\begin{equation}
  M = {67.3}_{-42.5}^{+103.2}~M_{\oplus}, \quad D_{\rm L} = {7.34}_{-2.11}^{+0.96}~{\rm kpc}.   
  \label{bayesian}
\end{equation}
Our Bayesian results suggest that the lens would be a Saturn-like planet located in the bulge.

As with other FFP candidate events, we cannot rule out the existence of a widely separated host star. We applied 2L1S models to the event to investigate possible signals for the host, but found that the fit improvement by the best-fit 2L1S model is only $\Delta\chi^{2} = 6.4$. For typical events with $M_{\rm host} = 0.6\,M_{\odot}$ and $D_{\rm L} = 4.5\,{\rm kpc}$, we detected no significant evidence for the lens host within the planet-host projected distance of $a_{\bot} \sim 6\,{\rm au}$.

If the lens hosts a star, the host can be identified through high-resolution follow-up imaging. For KMT-2024-BLG-3237, however, the detection probability for possible hosts with current telescopes would be very low, due to the relatively bright source of $K_{\rm S} \simeq 13.02$. Recently, \citet{Mroz2024} carried out follow-up observations using Keck adaptive optics, together with sensitivity simulations for putative hosts, and calculated the detection probability of possible hosts for the six PSPL FFP candidates found by \citet{Mroz2017}. Based on their results, \citet{Jung2024} estimated that for KMT-2023-BLG-2669, the Keck host detection probability would be $p \sim 17\%$ at angular lens-source separations of $\Delta\theta \sim 200\,{\rm mas}$. Given that our source is similar to that of \citet{Jung2024}, we follow the approach used for KMT-2023-BLG-2669, and estimate the detection probability corresponding to our event as $p \sim 16\%$ at $\Delta\theta \sim 200\,{\rm mas}$. This suggests that it is difficult to make a distinction between wide-orbit and free-floating planet hypotheses with current telescopes. Nevertheless, because KMT-2024-BLG-3237 is an FFP candidate with the measurement of $\mu_{\rm rel}$, it would be possible to carry out a search for possible hosts within a few years after the event \citep{Jung2024}.

Finally, we note that if the upcoming extremely large telescopes (ELTs) achieve the contrast ratio between the source and the host of ${\Delta}K \sim 5\,{\rm mag}$ at $\Delta\theta = 200\,(D/10)^{-1}\,{\rm mas}$, a putative Sun-like host could be precisely identified in ELTs images taken before the end of the next decade. Here, $D$ is the mirror diameter. For a putative host with ${\Delta}K \sim 11\,{\rm mag}$ ($M$-type dwarf)\footnote{Hosts have not been identified in the presence of such high contrast ratios, and it is extremely difficult to identify them with current telescopes (see Figure 1 of \citealt{Mroz2024}).}, the wait time will be longer, but it is difficult to make an estimate for this wait time without measuring the performance of ELTs.

\begin{acknowledgments}
% Collaboration acknowledgement
This research has made use of the KMTNet system operated by the Korea Astronomy and Space Science Institute (KASI) at three host sites of CTIO in Chile, SAAO in South Africa, and SSO in Australia. Data transfer from the host site to KASI was supported by the Korea Research Environment Open NETwork (KREONET). 
% KASI research project acknowledgement
This research was supported by KASI under the R\&D program (Project No. 2025-1-830-05) supervised by the Ministry of Science and ICT.
% Individual acknowledgements
H.Y. and W.Z. acknowledge support by the National Natural Science Foundation of China (Grant No. 12133005). 
H.Y. acknowledge support by the China Postdoctoral Science Foundation (No. 2024M762938).
Work by C.H. was supported by the grants of National Research Foundation of Korea (2019R1A2C2085965 and 2020R1A4A2002885). 
\end{acknowledgments}

\clearpage

\begin{deluxetable}{lrr}
\tablecaption{Mean Parameters for FSPL Models}
\tablewidth{0pt}
\tablehead{
\multicolumn{1}{l}{Parameters} &
\multicolumn{1}{c}{FSPL} &
\multicolumn{1}{c}{FSPL $(f_{\rm B} = 0)$}
}
\startdata
$\chi^2$/dof                 &   6006.678/6005         &     6007.374/6006  \\
$t_0$ $(\rm{HJD}^{\prime})$  & 541.902 $\pm$ 0.003     & 541.902 $\pm$ 0.003\\  
$u_0$                        &   0.436 $\pm$ 0.301     &   0.465 $\pm$ 0.042\\  
$t_{\rm E}$ (days)           &   0.561 $\pm$ 0.088     &   0.536 $\pm$ 0.018\\  
$\rho$                       &   1.008 $\pm$ 0.265     &   1.018 $\pm$ 0.030\\  
$f_{\rm S}$ (KMTS41)         &   0.819 $\pm$ 0.410     &   0.785 $\pm$ 0.0001\\  
$f_{\rm B}$ (KMTS41)         &  -0.033 $\pm$ 0.410     &         ...  
\enddata
\tablecomments{$\rm{HJD}^{\prime} = \rm{HJD} - 2460000$ 
}
\label{tab:FSPL}
\end{deluxetable}
    
\begin{deluxetable}{lrrr}
\tablecaption{Mean Parameters for 2L1S Models}
\tablewidth{0pt}
\tablehead{
\colhead{Parameter} & 
\colhead{Local A} & 
\multicolumn{2}{c}{Local B} \\
\colhead{} & \colhead{} & \colhead{Free $f_{\rm B}$} & \colhead{$f_{\rm B}=0$}
}
\startdata
$\chi^2/\mathrm{dof}$     & 6006.237/6002              & 6000.963/6002              & 6001.201/6003 \\
$t_0$ (HJD$^\prime$)    & $542.041 \pm 0.360$          & $541.901 \pm 0.004$        & $541.902 \pm 0.004$ \\
$u_0$                     & $0.386 \pm 0.253$          &   $0.309 \pm 0.173$        &   $0.274 \pm 0.048$ \\
$t_{\rm E}$ (days)        & $0.598 \pm 0.074$          &   $0.944 \pm 0.168$        &   $0.929 \pm 0.152$ \\
$s$                       & $1.315 \pm 0.351$          &  $27.987 \pm 23.516$       &  $30.005 \pm 23.727$ \\
$q$ ($10^2$)       & $708.051_{-36.433}^{+801.884}$    &   $0.026 \pm 0.021$        &   $0.021 \pm 0.011$ \\
$\alpha$                 & $-0.438 \pm 3.241$          &  $-0.864 \pm 3.437$        &   $0.066 \pm 3.324$ \\
$\rho$                    & $0.895 \pm 0.175$          &   $0.624 \pm 0.148$        &   $0.601 \pm 0.084$ \\
$f_{\rm S}$ (KMTS41)      & $0.661 \pm 0.273$          &   $1.081 \pm 0.829$        &   $0.785 \pm 0.0001$ \\
$f_{\rm B}$ (KMTS41)      & $0.124 \pm 0.274$          &  $-0.296 \pm 0.829$        &     $\dots$ \\
\enddata
\tablecomments{HJD$^\prime \equiv$ HJD $-\,2460000$}
\label{tab:2L1S}
\end{deluxetable}

\clearpage

\clearpage
\bibliographystyle{aasjournal}
\bibliography{references}

\end{document}